\documentstyle[prl,aps,multicol,epsf]{revtex}

\newcommand{\be}[1]{\begin{equation}\label{#1}}%
\newcommand{\ee}{\end{equation}}%
\newcommand{\bea}[1]{\begin{eqnarray}\label{#1}}%
\newcommand{\eea}{\end{eqnarray}}%
\newcommand{\ba}[1]{\left(\begin{array}{#1}}
\newcommand{\ea}{\end{array}\right)}%
\newcommand{\hide}[1]{}
\newcommand{\gtopl}{\mbox{\hbox{ \lower-.6ex\hbox{$>$}\kern-.8em
\lower.5ex\hbox{$<$}\kern+.35em}}}
\newcommand{\ltopg}{\mbox{\hbox{ \lower-.6ex\hbox{$<$}\kern-.8em
\lower.5ex\hbox{$>$}\kern+.35em}}}
\newcounter{saveeqn}%
\newcommand{\alpheqn}[1]{\addtocounter{equation}{1}
{\immediate\write1{\string\newlabel{#1}{{\theequation}{\thepage}}}}
\setcounter{saveeqn}{\value{equation}}\setcounter{equation}{0}
\renewcommand{\theequation}{\mbox{\arabic{saveeqn}\alph{equation}}}}
\newcommand{\reseteqn}{\setcounter{equation}{\value{saveeqn}}
\renewcommand{\theequation}{\arabic{equation}}}

\begin{document}
\draft
\title{Mobility in a one-dimensional disorder potential
\footnote{Dedicated to Professor H. Wagner
on the occasion of his 60th birthday}}
\author{Stefan Scheidl
\footnote{Address after Oct. 1, 1994: Inst. f\"ur Theoret. Physik,
Universit\"at zu K\"oln, Z\"ulpicher Str. 77, 50937 K\"oln, Germany}}
\address{Institut Laue-Langevin, BP 156, 38042 Grenoble, France \\
e-mail: scheidl@gaucho.ill.fr}

\date{}
\maketitle

\begin{abstract}
In this article the one-dimensional, overdamped motion of a classical
particle is considered, which is coupled to a thermal bath and is
drifting in a quenched disorder potential. The mobility of the
particle is examined as a function of temperature and driving force
acting on the particle. A framework is presented, which reveals the
dependence of mobility on spatial correlations of the disorder
potential. Mobility is then calculated explicitly for new models of
disorder, in particular with spatial correlations. It exhibits
interesting dynamical phenomena. Most markedly, the temperature
dependence of mobility may deviate qualitatively from Arrhenius
formula and a localization transition from zero to finite mobility may
occur at finite temperature. Examples show a suppression of this
transition by disorder correlations.
\end{abstract}

\begin{multicols}{2}
\narrowtext

\newcommand{\figone}{
\begin{figure}[b]
\vskip 6.7cm
\hskip -0.7cm
\epsffile{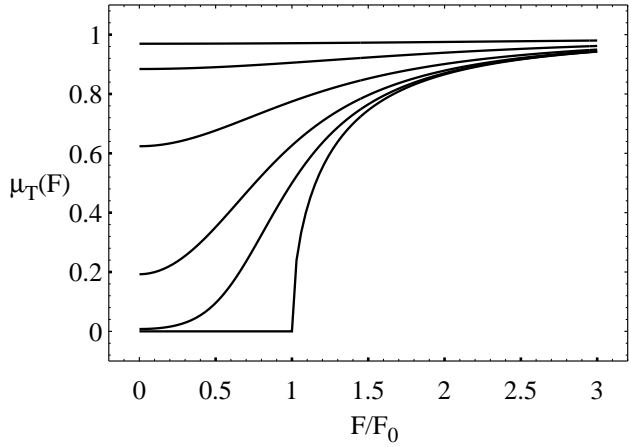}
\caption{Mobility of the sinusoidal model according to
Eq.~(\protect\ref{v.sin.exact}) as a function of the driving force for
$T/E=4$, 2, 1, 0.5, 0.25, 0 from top to bottom.}
\label{capt_sine}
\end{figure}
}
\newcommand{\figtwo}{
\begin{figure}[b]
\vskip 17.1cm
\hskip -0.7cm
\epsffile{fig2.epsf}
\caption{Mobility of the Gaussian model as a function of the driving force
for different temperatures $T=4$, 2, 1, 0.5, 0.25, 0.125 from top to
bottom in each diagram: (a) for $\protect m_0=m_2=1$ according to
Eq.~(\protect\ref{v.gauss2}), (b) for $\protect m_1=m_2=1$ after
numerical Laplace integration of Eq.~(\protect\ref{g.gauss1}), and (c)
for the random force model as limiting case $\protect m_0=m_1=0,
m_2=1$.}
\label{capt_Gauss}
\end{figure}
}
\newcommand{\figthree}{
\begin{figure}[b]
\vskip 17.5cm
\hskip -0.7cm
\epsffile{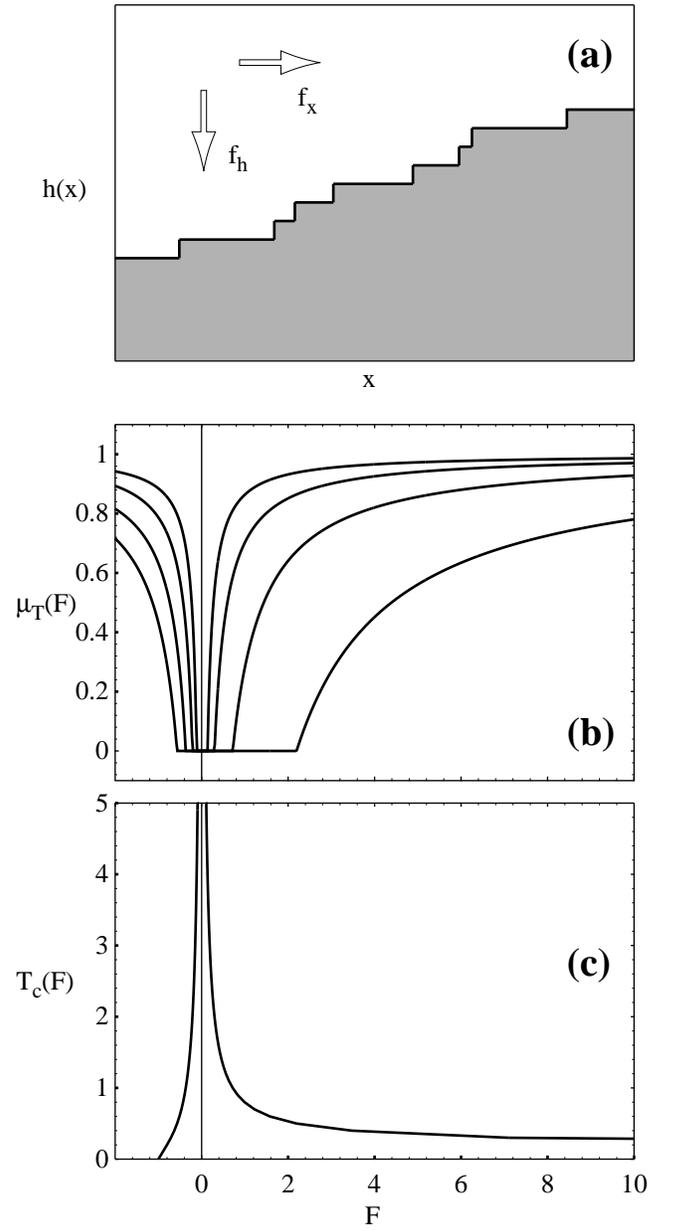}
\caption{Poissonian model: (a) sketch of geometry, (b) mobility according to
Eq.~(\protect\ref{v.Poisson}) as a function of the driving force at
temperatures $T=4$, 2, 1, 0.5 from top to bottom and (c) localization
temperature as a function of driving force for $a f_h=l_0=1$.}
\label{capt_Poisson}
\end{figure}
}

\section{Introduction}

In recent years there has been wide interest in transport properties
of disordered media: e.g. diffusion on a polymer in an external field,
random resistor networks, domain wall dynamics of magnets in a random
field, and pinning of vortices in type-II superconductors. Some
reviews have already been devoted to this subject
\cite{Ale+81,HK87,HB87,BG90}.

This article focuses on the mobility of a particle moving at finite
temperature in a one-dimensional disorder potential. Its purpose is
twofold: (i) The functional dependence of mobility on temperature and
on an additional external driving force is for the first time worked
out explicitly in terms of stochastic properties of the disorder
potential in the continuous space. Previous publications have mainly
focused on the situation, where the particle moves on a lattice. The
dynamics was specified in terms of hopping rates between neighbored
places. Our point of view will be more adequate and also physically
more transparent in situations, where the disorder potential is well
characterized and hopping rates, if to be used, had to be calculated
from this potential first. In other works, where space was treated as
continuous, only special disorder types have been considered
\cite{Bou+90,Asl+91}. (ii) This functional dependence is explicitly
calculated and discussed for new models, in particular for spatially
correlated distributions of the disorder force. Thereby we obtain
generalizations of the Sinai model \cite{Sinai} with spatially
uncorrelated forces, which has attracted particular attention in the
past (see e.g. \cite{Bou+90} and references therein).

Our approach, which leads to closed analytic expressions, is limited
to the evaluation of the mean velocity. Thus interesting transport
phenomena, like an anomalous scaling behavior of the (mean squared)
displacement as a function of time, which characterize dynamical
phases \cite{BG90} and were found for the Sinai model
\cite{Bou+90,Asl+91,Sinai,Derrida,Bou+87EPL}, are beyond the scope of
the present treatment.

In the following, we first derive and discuss the general expression
for mobility (Sec.~\ref{sec_2}). Its asymptotic behavior for large or
small temperatures or driving forces is then analyzed
(Sec.~\ref{sec_3}). In Sec.~\ref{sec_4} mobility is calculated over
the complete parameter range for some models. Finally, the phenomena
of thermal activation encountered thereby are summarized
(Sec.~\ref{sec_5}).

\section{Basic description}
\label{sec_2}

Our problem is defined by the one-dimensional Langevin-equation for a
single particle with coordinate $x$ in the presence of a disorder
potential $U(x)$ and an external force $F$,
\be{Langevin}
  \dot{x}(t) = F - U'(x(t)) + \eta(t) .
\ee

The thermal random force $\eta(t)$ is assumed to be Gaussian
distributed with moments
\alpheqn{zeta.mom}
\bea{zeta.mom.a}
  \langle \eta(t) \rangle &=& 0 , \\
  \langle \eta(t) \eta(t') \rangle &=& 2 T \delta(t-t') .
\eea
\reseteqn
Angular brackets represent thermal average in a heat bath of
temperature $T$.

We are interested in the velocity-force-characteristics (VFC) for a
{\em given} disorder potential, where the average velocity is defined
by
\be{def.v}
  v_T(F) := \lim_{t \rightarrow \infty} \frac 1t \
  \overline{ \langle x(t)-x(0) \rangle} .
\ee
A bar denotes the translation-average for the {\em given} realization
of disorder, i.e. one replaces $U(x)$ by $U(x-x_0)$ and averages over
all $x_0$.  In the last formula, this means an average over all
initial positions $x(0)$.  If desired, an additional average over an
ensemble of disorder realizations should be taken as last operation.

For convenience, we suppose for the moment periodicity of the
potential, $U(x+L)=U(x)$, and formulate the result in a way, which
does not depend on the periodicity $L$. We argue then, that the result
is correct for any unbiased disorder potential.

The dynamics of the model can be reformulated in terms of the
Fok\-ker-Planck equation for the probability density $P(t,x)$ and
current density $J(t,x)$:
\alpheqn{FPE}
\bea{FPE.a}
  \partial_t P(t,x) &=& - \ \partial_x J(t,x) ,\\
  J(t,x) &=& \left[ F-U'(x) \right] P(t,x) - T \ \partial_x P(t,x)  .
\eea
\reseteqn
 From the stationary solution, which has a homogeneous current
distribution, one derives the mean velocity $v=JL/\int_0^L dx P(x)$ in
a standard way (see e.g.
\cite{Ris84})
\bea{v.L}
  v_T(F) &=& TL \left(1- e^{-LF/T}\right) / \nonumber \\
  & & / \Bigg\{
  \int_0^L dx \ e^{- \phi(x)/T} \int_x^L dx' \ e^{\phi(x')/T} +
  \nonumber \\
  & & + e^{-LF/T} \int_0^L dx \ e^{-\phi(x)/T}
  \int_0^x dx' \ e^{\phi(x')/T} \Bigg\}
\eea
with the effective potential $\phi(x):=U(x) - F x$. This expression
simplifies by replacing the periodicity $L$ by $NL$ and taking the
limit $N\rightarrow \infty$. The result can be represented in two
ways. A first one, frequently used in the literature, will be
indicated in the following paragraph. The second one, which seemingly
has been disregarded up to now and which can be evaluated more easily,
is the basis of the remainder of this article.

The first version has the compact form
\alpheqn{VFC.1}
\be{v.1}
  v_T(F)=  \left[\  {\overline{ \tau(x)}} \ \right]^{-1} ,
\ee
where
\be{tau}
  \tau(x) := T^{-1} \int_x^{\pm \infty} dx' \ e^{[\phi(x')-\phi(x)]/T}
\ee
\reseteqn
with the integral running towards $\pm \infty$ for $F \gtopl 0$. This
representation of the VFC evaluates the so-called {\em sojourn-time
density} $\tau(x)$: $dt=\tau(x) dx$ is proportional to the conditional
probability of finding the particle in the interval $(x,x+dx)$ at
times $t>0$, provided it started from position $x$ at $t=0$. Therein
all later passages of the particle parallel and opposite to the
direction of the driving force are included.  Several works have been
devoted to the study of sojourn-time distributions, mainly in
spatially discrete hopping models (see e.g. \cite{Bou+90} and
references therein).

In contrast to Eq.~(\ref{VFC.1}), which emphasizes the ``dynamical''
aspect of the problem by the statistical analysis of the sojourn-time,
we prefer a second ``static'' point of view, formulated in terms of
correlations of the disorder potential. This formulation enables us
also to interpret our results directly in terms of activation
processes in the energy landscape of the disorder potential. It is
related to Eqs.~(\ref{VFC.1}) by a mere change of the order of two
integrations: the mobility
\be{def.mob}
  \mu_T(F):=v_T(F)/F
\ee
can be calculated directly as
\alpheqn{VFC.2}
\be{v.2}
  \mu_T(F)= \left\{ \int_{0}^{\infty} d \xi \ e^{-\xi}
  \ g_T(\xi T/F) \right\}^{-1}
\ee
with the generating function
\be{def.g}
  g_T(y) := \overline {\exp \left\{[U(x+y)-U(x)]/T \right\} }.
\ee
\reseteqn
Viewed as a function of temperature, $g_T(y)$ is the generating
function of the potential energy difference correlations at distance
$y$. Eq.~(\ref{v.2}) shows, that mobility is essentially the Laplace
transform of the generating function. In an experimental situation,
where the force- and temperature dependence of mobility are known, the
generating function may thus be determined by an inverse Laplace
transformation.

The above derivation was based on the fact, that the current density
is spatially constant in the stationary state, which is a
particularity of one-dimensional problems. Unfortunately this approach
does not allow an evaluation of the diffusion constant, which requires
joint probability distributions at different times, and relaxation
phenomena, which are not stationary.  Both topics have been treated
for the Sinai model\cite{Bou+90,Asl+91,Sinai,Derrida,Bou+87EPL}.

Formulae (\ref{VFC.1}) and (\ref{VFC.2}) have been derived for
periodic potentials. Since periodicity shows up only implicitly as a
property of $\tau(x)$ and $g_T(y)$ but no longer explicitly as
parameter, we postulate their validity also for non-periodic disorder
potentials. Depending on the nature of disorder it may happen, that
the stationary current is zero for some range of forces. In this case
the integral in Eq. (\ref{v.2}) will diverge, leading to vanishing
mobility.

In contrast to the original equation of motion, the expressions for
the mobility are not invariant under the transformation $U(x)
\rightarrow U(x) + const \times x$ and $F
\rightarrow F+ const$. They require an unbiased disorder potential,
i.e.
\be{unbias}
  \overline{ U'(x) }=0 .
\ee
This condition, which is obvious for periodic potentials, has to be
imposed on the non-periodic case as well. However, situations with a
biased disorder potential can be treated, too. Then in our expressions
$U$ has to be taken as the original potential after subtraction of its
bias and $F$ as the original external force plus the mean force of the
original potential.

For the spatially discrete version of the model with uncorrelated
hopping rates, it was shown\cite{Asl+89} that the results, obtained
from a periodic potential and taking the limit of infinite periodicity
in the end\cite{Derrida,Asl+89JPF}, are unchanged, if one allows for
non-periodicity from the very beginning. In addition, for zero
temperature but arbitrary disorder, one can integrate the equation of
motion Eq.~(\ref{Langevin}) after separation of variables and finds
directly the zero temperature limit of Eq.~(\ref{v.2}) without use of
periodicity, leading to Eq.~(\ref{v.T=0}) below.

Now we address the question, whether an additional {\rm average over
different realizations of the disorder potential} may affect the VFC.
In principle, this average should be taken as last average in
Eqs.~(\ref{v.1}) or (\ref{v.2}). Again, for the spatially discrete
version of the model with uncorrelated hopping rates, it was
shown\cite{Asl+89}, that velocity is a self-averaging quantity. This
means, that its value, when calculated for a given realization of the
disorder potential, coincides with probability one with the value
after an additional averaging over all realizations of the disorder
potential. It seems natural to assume this property for any disorder
potential with short-ranged correlations, since the particle samples
during its drift the potential on infinite length scales, where a
given realization is expected to be representative for the whole
ensemble of realizations. Therefore we may consider the spatial
averages as averages in the ensemble of disorder realizations. If, for
a contrary example, the disorder extends only over a finite region,
the ensemble-average does modify the result and is to be performed in
addition. This situation has been studied for the Sinai model without
bias\cite{OMM93}.

Our main formula (\ref{VFC.2}) can be illustrated physically in the
following way: Consider jumps over a finite distance $y$. The larger
the fluctuations of $U(x+y)-U(x)$ when $x$ varies, the larger will be
$g_T(y) \geq 1$. The inequality holds for all unbiased potentials.
Therefore we might call $\Delta_T(y):=T^{-1} \ln g_T(y) \geq 0$
``energetic roughness'' on distances $y$. A large roughness signifies
a pronounced relief-structure of the potential to be overcome by
thermal activation and reduces mobility. An ``enthalpic roughness''
$\hat \Delta_T(F)$ can be introduced by $\exp[\hat \Delta_T(F)/T]:=
(T/F) \int_0^{\pm \infty} dy \exp[-yF/T + \Delta_T(y)/T]$,
reminding of the relation between free energy and enthalpy in
equilibrium thermodynamics, since $F$ and $y$ are thermodynamically
conjugated variables. This enthalpic roughness turns out to act as an
effective activation energy, since it determines mobility by
$\mu_T(F)=\exp[-\hat \Delta_T(F)/T]$.

Eq.~(\ref{VFC.2}) shows also, that in the evaluation of the disorder
potential an energy scale is set by temperature, whereas the spatial
structure of the potential is relevant only up to the length scale
$T/F$.

\section{Limiting cases}
\label{sec_3}

Before we attempt to evaluate the VFC for particular models, defined
by a probability distribution of the disorder potential, we analyze
different limits of the general expression for mobility.

\subsection{Low temperatures}

At strictly zero temperature, we obtain from an integration of the
equation of motion (\ref{Langevin}) over a finite time interval
$(t_i,t_f)$:
\be{determinist}
  \frac{x_f-x_i}{t_f-t_i}= F \ \frac{x_f-x_i}{\int_{x_i}^{x_f}
  \frac{dx}{1-U'(x)/F}},
\ee
with $x_{i,f}=x(t_{i,f})$. If the disorder force is everywhere weaker
than the external force, $U'(x)/F<1$, the particle cannot be trapped
in the disorder potential. In the limit $t_f-t_i \rightarrow \infty$,
Eq.~(\ref{determinist}) then confirms the zero-temperature limit of
Eq.~(\ref{VFC.2}):
\be{v.T=0}
  \mu_T(F)= \left\{ \overline{[1-U'(x)/F]^{-1}}\right\}^{-1}.
\ee
In the opposite case, with $U'(x)/F>1$ somewhere, the particle will be
localized, i.e. have vanishing mobility. The threshold-forces
$F_{c}^\pm$ for the onset of drift towards $x= \pm \infty$ are clearly
given by the maximum/minimum slope of $U$. In the localized region,
Eq.~(\ref{determinist}) is invalid in the strict sense, since the
particle does no longer sample the whole potential. However, if it is
used naively, localization formally shows up by a divergence of the
integral in the denominator.

At small, but finite temperatures, one might expect to find an
Arrhenius-like thermally activated behavior. This is certainly true
for periodic potentials, where one easily derives (for $F \ltopg
F_{c}^\pm$)
\be{v.kram}
  v=L \frac{\sqrt{-\lambda_{\max} \lambda_{\min}}}{2 \pi}
  \ e^{-[\phi(x_{\max})-\phi(x_{\min})]/T}
\ee
with $\lambda_{\max,\min}:= \phi''(x_{\max,\min})$, where
$x_{\min/\max}$ denotes the position of a minimum/maximum of $\phi$
such, that $(x_{\max}-x_{\min})/F>0$ and the energy difference
$\phi(x_{\max})-\phi(x_{\min})$ is maximal. This expression is valid
only for temperatures much smaller than this activation energy and
$L|F|/T \gg 1$, such that activation over the maximum occurs only in
the direction of $F$. In this case, the mean velocity is just
proportional to Kramers transition rate (see e.g. \cite{HTB90}) for
thermal activation out of minimum $x_{\min}$ over $x_{\max}$. The
resulting velocity is finite for all finite external forces.

In the case of non-periodic disorder potentials, where arbitrarily
large energy barriers (or curvatures at extrema) occur with finite
probability density and thus no highest energy barrier exists,
deviations from Arrhenius-like temperature dependence may occur.

\subsection{High temperatures}

For high temperatures one may expand the generating function into
\be{g.calc}
  g_T(y)=\sum_{n=0}^{\infty} \frac{1}{n!} T^{-n} \
  \overline{[U(x+y)-U(x)]^n}.
\ee
An additional temperature dependence of mobility comes in through the
fact, that the generating function is evaluated at distances $y=\xi
T/F$, which increase with temperature. If qualitatively
$\overline{[U(x+y)-U(x)]^n} \sim |y|^{n \zeta}$ for $|y| \rightarrow
\infty$ with some roughness exponent $\zeta < 1$, we will have anyway
\be{v.T.infty}
  \mu_T(F) = 1 - {\cal O}\left(T^{2( \zeta -1)}\right) .
\ee
On the other hand, if the potential is so rough that $\zeta > 1$, this
series diverges and the particle can be expected to be localized at
all temperatures.

\subsection{Small driving forces}

In the limit of small $F$, the function $g_T(y)$ is evaluated in
(\ref{v.2}) mainly for large arguments and the integral therein acts
as additional spatial average. Then one has
\be{asy.v.F.0}
  \mu_T(F) \approx \left[ \overline {e^{U(x)/T}} \ \
  \overline {e^{-U(x)/T}} \right]^{-1},
\ee
provided that the averages therein exist. Therefore, one has finite
mobility $0<\mu \leq 1$ for small driving forces, i.e. Ohmic behavior.
The range of this Ohmic regime is given by the condition, that $T/F$
has to be small compared to typical length scales (periodicity or
correlation length) of the disorder. Remarkably, mobility is
independent of the direction of the driving force, even for asymmetric
potentials. If, on the other hand, one of the averages is infinite,
mobility vanishes for small forces. In general, any non-Ohmic behavior
requires the divergence of one of the averages in
Eq.~(\ref{asy.v.F.0}).

It is interesting to note, that a disorder potential may give rise to
a temperature dependence of the Ohmic mobility, which is qualitatively
different from Arrhenius-like behavior. In order to illustrate this,
let us assume a potential energy density
\be{def.P(U)}
  P(U_1):=\overline{ \delta(U_1-U(x))}
\ee
with an asymptotics
\be{P(U).powerasy}
  P(U) \approx c \exp\left[-|U/E|^\sigma\right]
\ee
for $U\rightarrow \pm \infty$ with some energy scale $E$ and exponent
$\sigma>1$. For small temperatures we may evaluate the integrals in
Eq.~(\ref{asy.v.F.0}) with a saddle-point approximation and obtain
\bea{v.power}
  \mu_T(F) &\approx& \frac{\sigma (\sigma -1)}{2 \pi c^2 E^2}
  \left( \frac{E}{\sigma T}\right)^{(\sigma-2)/(\sigma-1)} \times \nonumber
  \\
  & & \times \exp \left[ -2 (\sigma-1) \left(
  \frac{E}{\sigma T}\right)^{\sigma/(\sigma-1)} \right] .
\eea
Only in the limit $\sigma \rightarrow \infty$, i.e. a narrow
distribution $P(U)$, one obtains an Arrhenius-like temperature
dependence of the velocity.  Potentials of a finite width, like
periodic or quasiperiodic ones, belong to this class. For general
$\sigma>1$, mobility will vanish with temperature faster than
Arrhenius-like, since the exponent $\sigma/(\sigma-1)$ in the
exponential function will be greater than one. However, for $\sigma>1$
and $T>0$ mobility will always be finite.

In the marginal case $\sigma=1$ we find a transition from zero
mobility at $T \leq E$ to finite mobility at $T>E$. In the case
$\sigma<1$ the particle will always have zero mobility.

These results show the naively expected tendency: the broader the
distribution $P(U)$, the larger are the involved activation energies
and the smaller is the mobility. Perhaps less expected is the fact,
that spatial correlations of the potential for small $y$ will affect
the VFC only for larger driving forces.

\subsection{Large driving forces}

In the limit of large $F$, the function $g_T(y)$ is evaluated in
Eq.~(\ref{v.2}) only for small arguments. We now expand
\be{asy.g.xi=0}
  g_T(\xi T/F) = 1+ (F_{c1}^\pm /F) \ \xi +
  \frac{1}{2} (F_{c2}^\pm/ F^2) \ \xi^2 + \dots
\ee
for $\xi \gtopl 0$, where we introduced in general temperature-dependent
characteristic forces of order $n$ by
\be{def.Fcn}
  F_{cn}^\pm:=\left.\left( T \frac{d}{d y}\right)^n
  g_T(y) \right|_{y=0^\pm} .
\ee
 From the representation (\ref{v.2}) for the mean velocity, we obtain
\be{asy.v.F.infty}
\mu_T(F) \approx \frac{1}{ 1+F_{c1}^\pm/F+F_{c2}^\pm/F^2+ \dots}
\ee

If the nature of the disorder potential is ``smooth'', i.e. allows to
change the order of differentiation an averaging, one derives from the
definition (\ref{def.g}) of $g$, that
\alpheqn{Fcn.smooth}
\bea{Fcn.smooth.a}
F_{c1}^\pm &=& \overline{ U'(x) }=0 , \\
F_{c2}^\pm &=& \overline{ T U''(x) + U'^2(x) }>0 .
\eea
\reseteqn
If in addition $\overline{U''(x)}=0$, as for periodic potentials, the
leading correction of the asymptotics is temperature-independent.

Otherwise, if the potential is not smooth, the identities
(\ref{Fcn.smooth}) need not hold, as we will see in the Gaussian model
below, where $F_{c1}^\pm$ may be different from zero. Then disorder
affects a finite shift of velocity with respect to the disorder-free
case even for large velocities,
\be{v.shift}
  v_T(F \rightarrow \pm \infty) = F - F_{c1}^\pm +{\cal O}(F^{-1})
\ee
for $F^2 \gg F_{c2}^\pm$.

In the case of a discontinuous potential, one may find $g_T(0^\pm)
\neq 1$ inhibiting the expansion (\ref{asy.g.xi=0}), see e.g. the
random-potential case of the Gaussian model below.


\section{Examples}
\label{sec_4}

We now explicitly calculate the VFC for certain disorder potentials to
demonstrate the richness of phenomena which can be found in this class
of diffusion problems.

\subsection{Sinusoidal model}

For the sake of completeness, let us discuss in the present context
the periodic model $U(x)=E\sin(2 \pi x/L)$, which has been studied
already some time ago by Ambegaokar and Halperin\cite{AH69} in order
to determine the thermal noise contribution to the dc Josephson
effect. Afterwards, we will discuss a non-periodic generalization of
this model.

The generating function can easily be calculated as
\be{g.sin}
  g_T(y)= I_0 \left[2 (E/T) \sin(\pi y/L) \right] ,
\ee
where $I_0$ is a modified Bessel function. Due to the smoothness and
symmetry of the potential, $g_T(y)$ is even and analytic. A
straightforward evaluation of Eq.~(\ref{v.2}) leads to
\bea{v.sin.exact}
  \mu_T(F) &=& \frac{\sinh \pi f/\vartheta}{\pi f/\vartheta} \
  \Bigg\{ \int_0^1 d \xi \times \nonumber \\
  & & \times \cosh [\xi \pi f/\vartheta] \ I_0 [(2/\vartheta) \cos (\xi \pi/2)]
  \Bigg\}^{-1}
\eea
with reduced temperature $\vartheta:= T/E$ and reduced force
$f:=F/F_0$ where $F_0:=2 \pi E/L$. This expression can be treated
further analytically only in limiting cases. The result of its
numerical integration is shown in Fig.~\ref{capt_sine}.
\hide{place_fig1}

At $T=0$ one finds from Eq.~(\ref{v.T=0})
\be{v.F.period.T=0}
  \mu_0(F)= \left\{  \begin{array}{ll}
     0 & {\rm for} \quad |f| \leq 1\\
     \sqrt{1-f^{-2}} &{\rm for} \quad |f|>1
  \end{array} \right.
\ee
with a rapid onset of motion near the the critical force
$F_c:=\max_x[|U'(x)|]=F_0$.

At low temperatures and within the force-region $|f|<1$, where the
particle is localized at zero temperature, we find from an saddle-point
evaluation of Eq.~(\ref{v.sin.exact})
\alpheqn{kramers.sin}
\bea{kramers.sin.a}
  v_T(F)&=& 2 F_0 \  \sinh(\pi f/\vartheta) \
            \sqrt{1-f^2} \ \ e^{-E_A(F)/T}  \\
  E_A(F)&:=&2 E \sqrt{1-f^2} +2 E f \arcsin f ,
\eea
\reseteqn
provided $\vartheta \ll \sqrt{1-f^2}$ and $\vartheta \ll f \arcsin f$.
The temperature range of its validity shrinks when the external force
vanishes or approaches the critical force. This asymptotic evaluation
yields the mean velocity just as the difference of Kramers rates for
an activation parallel and antiparallel to the driving force across
energy barriers $E_A(F) \mp \pi f E$.  Since the energy barriers
assume only certain values, temperature dependence is Arrhenius-like.

At high temperatures, we can expand the denominator of
Eq.~(\ref{v.sin.exact}) in $E/T$. This produces only even terms and we
find to second order
\be{sin.T.inf}
  \mu_T(F) \approx \left\{ 1+ \frac{2 (E/T)^2}{4+(2 f E/T)^2}
  \right\}^{-1} .
\ee

At small forces follows from Eq.~(\ref{v.sin.exact})
\be{sin.F.0}
  \mu_T(F) \approx [ I_0(E/T)]^{-2}
\ee
in agreement with a direct calculation from Eq.~(\ref{asy.v.F.0}) with
\be{sin.F.av}
  \overline{e^{\pm U(x)/T}}=I_0(E/T).
\ee

The characteristic forces of lowest order are $F_{c1}^\pm=0$ and
$F_{c2}^\pm= F_0^2/2$, since $U$ is smooth. In Fig.~\ref{capt_sine}
one observes for finite temperatures, that the VFC approaches in the
limit $F \rightarrow \infty$ first the curve $T=0$ before it assumes
the limit $\mu=1$. This reflects the fact, that the leading correction
to $\mu=1$ is temperature-independent.

Now we address the question, how these results will change when the
periodicity of the potential is abandoned. Let us construct such a
potential by stretching the maxima $U=E$ from points to plateaus such,
that the density of minima $U=-E$ is reduced to $\rho < 1/L$. The
generating function $g_T(y)$ behaves for small $y$ as
\be{g.xsin}
  g_T(y) \approx (1- \rho L) + \rho L I_0
  \left[2(E/T)\sin(\pi y/L) \right] ,
\ee
since this distance falls with probability $(1- \rho L)$ onto a
plateau and with probability $\rho L$ into a trap region with $U<E$,
where Eq.~(\ref{g.sin}) holds. This expression is correct in ${\cal
O}(y^2)$, thus the leading correction for large driving forces is
given by
\be{Fc2.xsin}
  F_{c2}^\pm= 2 \rho L (\pi E / L)^2
\ee
and vanishes with density $\rho$. However, the critical force $F_c=2
\pi E/L$ at $T=0$, being the maximum slope of $U$, does not depend on
$\rho$. On the other hand, if there is no long-range translational
order between the traps, we deduce using Eq.~(\ref{sin.F.av})
\be{xsin.F.av}
  \overline{e^{\pm U(x)/T}} \approx
  (1- \rho L) e^{\pm E/T} + \rho L I_0(E/T).
\ee
and the asymptotics
\bea{xsin.asy}
  \mu_T(F \rightarrow 0) &\approx& \big[
  2(1- \rho L)\rho L \sinh(E/T) I_0(E/T) +  \nonumber \\
  & & + (1- \rho L)^2 +\rho^2 L^2 I_0^2(E/T) \big]^{-1}.
\eea
This expression displays a low-temperature behavior
\bea{xsin.asy.lt}
  \mu_T(F \rightarrow 0) &\approx&
  \left[ (1- \rho L)\rho L \sqrt{2 \pi E/T} + \rho^2 L^2 \right]^{-1}
  \times \nonumber \\
  & & \times 2\pi (E/T) e^{-2E/T}.
\eea
Its exponential factor is Arrhenius-like with activation energy $2E$.
The prefactor can {\em not} be calculated from Kramers expression
(\ref{v.kram}), since the curvature of the maximum is infinite.
Remarkably, mobility at zero force and its leading correction at large
forces depend only on the mean density of traps and not on more
information about their distribution. This is true for any shape of
the traps, at small forces since $\overline{\exp[\pm U(x)/T]}$ depends
only on the shape and density of traps, at large forces since the
leading term of $g_T(y)-1$ for small $y$ is simply proportional to the
density.

\subsection{Gaussian model}

In general, the probability distribution ${\cal P}$ of the functions
$U$ may be generated by an ``hamiltonian'' ${\cal H}$,
\be{P[U].ham}
  {\cal P}[U] \propto \exp\{-{\cal H}[U]\}.
\ee
The hamiltonian can have direct physical significance. Imagine that
the particle diffuses on a substrate. In an external field, the shape
of this substrate will determine the energy $U$ of the particle.
Assume that the shape of the substrate was frozen after a quench from
an initial temperature, where the substrate performed
shape-fluctuations according to a reduced hamiltonian ${\cal H}$
(substrate energy divided by temperature). Then this hamiltonian
determines the distribution of potentials $U$ according to
Eq.~(\ref{P[U].ham}).

In this section we consider the Gaussian case, where the hamiltonian
is bilinear in the potential. We assume the ``energy'' of a
realization of the potential to be determined by a inversion-symmetric
stiffness $\epsilon$ in Fourier-space according to
\be{H.gauss}
  {\cal H}[U] = \frac 12 \int \frac{dk}{2 \pi} \ U(k) \epsilon (k) U(-k),
\ee
for which one obtains immediately
\be{g.gauss}
  g_T(y)=\exp \left\{ \frac1{T^2} \int \frac{dk}{2 \pi} \
  \frac {1-\cos(k y)}{\epsilon(k)} \right\} .
\ee
The stiffness should satisfy $\lim_{k \rightarrow 0} k^3/\epsilon(k)
=0= \lim_{k \rightarrow \infty} k/\epsilon(k)$, otherwise the particle
is always localized. We restrict us to $\epsilon(k)=m_0+m_1|k|+m_2
k^2$, thereby generalizing Sinai's random-force model with a term $m_2
k^2$ only.  For simplification we select the cases with $m_1=0$ or
$m_0=0$.

We start with $m_1=0$, where the absolute fluctuations of $U(x)$ are
confined.  One has
\be{g.gauss2}
  g_T(y)=\exp \left\{\left(1-e^{-|y|/y_0}\right) y_0 F_T/T  \right\}
\ee
with the length- and force-scales
\alpheqn{F0.gauss2}
\bea{F0.gauss2.a}
  y_0&:=&\sqrt{m_2 /m_0} ,\\
  F_T&:=&1/ 2 m_2 T .
\eea
\reseteqn
Since the potential now typically is rough on short length scales,
$g_T(y)$ has a non-analytic distance dependence at $y=0$. In addition,
the first characteristic force
\be{Fc1.gauss2}
  F_{c1}^\pm=\pm F_T
\ee
is finite and depends on temperature, as well as the characteristic forces of
higher order.

Laplace integration over $g$ yields (see Fig.~\ref{capt_Gauss}a)
\be{v.gauss2}
  \mu_T(F) = \frac{1}{y_0 |F|/T}
  \frac{ (y_0 F_T/T)^{y_0 |F|/T} \ e^{-y_0 F_T/T} }
  {\gamma(y_0 |F|/T,y_0 F_T/T)} ,
\ee
\hide{place_fig2}
where $\gamma$ denotes the incomplete gamma function. This model
exhibits the asymptotic behavior
\alpheqn{v.gauss2.lim}
\bea{v.gauss2.lim.a}
  \mu_T(F \rightarrow \pm \infty) &=&1-F_T/|F| ,\\
  \mu_T(F \rightarrow 0) &=& \exp \{-y_0 F_T/T\} .
\eea
\reseteqn
It provides an explicit example of a non-Arrhenius-like temperature
dependence of the mobility, since $y_0 F_T/T=1/(2\sqrt{m_0 m_2} \
T^2)$. The fact, that temperature occurs squared, is due to the
Gaussian nature of disorder in agreement with the discussion leading
to Eq.~(\ref{v.power}). Note, that the limit $F \rightarrow 0$ always
contains the constant $m_2$, which enters the distribution
\alpheqn{P(U).Gauss}
\bea{P(U).Gauss.a}
P(U)&\propto& \exp \left\{-\frac {U^2} 2 \left[ \int \frac {dk}{2 \pi
\epsilon(k)} \right] ^{-1} \right\}=\\
&=&  \exp \{- \sqrt{m_0 m_2} \ U^2 \}.
\eea
\reseteqn
It would be therefore misleading to reason, that this limit should
depend only on $m_2$, which governs $\epsilon(k)$ for small $k$.
Rather, for this limit all length scales matter, as already visible in
Eq.~(\ref{g.gauss}).

The special case $m_1=m_2=0$ describes an uncorrelated potential and
Eq.~(\ref{asy.v.F.0}) applies for all forces. As the amplitude of
fluctuations diverges in the limit $m_2 \rightarrow 0$, the generating
function diverges since $\lim_{k \rightarrow \infty} k/\epsilon(k)
\neq 0$ and the particle is localized at all temperatures and driving
forces.

Consider now the case $m_0=0$, where
\bea{g.gauss1}
  g_T(y)&=&\exp \{ (2 y_1 F_T/\pi T) [C+\ln|y/y_1| - \nonumber \\
  & & - {\rm ci} \ |y/y_1| \cos |y/y_1| -
  {\rm si} \ |y/y_1| \sin |y/y_1| ] \}\\
  & \approx & \left( e^C |y/y_1| \right) ^{(2 y_1 F_T/\pi T)}
\label{g.gauss.power.2}
\eea
with a length-scale $y_1:=m_2/m_1$, the force-scale $F_T:=1/ 2 m_2 T$
as before, Eulers constant $C=0.577 \dots$, and the cosine and sine
integrals ${\rm ci}$ and ${\rm si}$. Fig.~\ref{capt_Gauss}b shows
mobility obtained by numerical integration of Eq.~(\ref{g.gauss1}).
The approximation in Eq.~(\ref{g.gauss.power.2}) refers to $|y/y_1|
\gg 1$. The power-law behavior of $g$ for large $y$ translates under
Laplace transformation to a power-law behavior for small $F$:
\bea{v.gauss1.power}
  \mu_T(F) &\approx& \frac
  {\left( e^{-C} y_1 |F|/T \right) ^{(2 y_1 F_T/\pi T)}}
  {\Gamma[(2 y_1 F_T/\pi T) +1]} \\
  &\rightarrow& \frac 1 {\sqrt{4 y_1 F_T/T}}
  \left( \pi e^{1-C}  |F| / 2 F_T \right) ^{(2 y_1 F_T/\pi T)}
\eea
$\Gamma$ denotes the gamma function, the approximation is valid for $F
\ll F_T$, the limit refers to $T \rightarrow 0$. Since the exponent $2
y_1 F_T/\pi T=1 / \pi m_1 T^2$ is positive, the mobility always
vanishes with $F$. The exponent is independent of the parameter $m_2$.
Mobility behaves at low temperatures like $T^{1/ \pi m_1 T^2}$. In
this case, we observe an even stranger deviation from
Arrhenius-activation. Since ${\cal P}[U]$ is invariant under $U(x)
\rightarrow U(x)+ const$, the potential is unbound and
Eq.~(\ref{v.power}) is violated.

The case $m_0=m_2=0$ again leads to complete localization, since
$\lim_{k \rightarrow \infty} k/\epsilon(k) \neq 0$.

The case $m_0=m_1=0$ is, as already indicated above, the Sinai
random-force model\cite{Sinai}. To show the equivalence of our
approach with the discrete approaches\cite{Derrida,Asl+89}, we
reproduce the VFC of this case. Eq.~(\ref{g.gauss2}) reduces to
\be{g.gauss2.0}
  g_T(y)=\exp \left\{|y| F_T/T \right\}
\ee
for all $y$ and results in (compare Fig.~\ref{capt_Gauss}c)
\be{v.gauss2.0}
  \mu_T(F)=\left\{ \begin{array}{ll}
  0 & {\rm for} \quad |F| \leq F_T \\
  1- F_T/|F| & {\rm for} \quad |F| > F_T .
\end{array} \right.
\ee
Therefore $F_T$ is the threshold force for the onset of motion. The
temperature dependence of the model is completely contained in this
threshold force. In particular, one finds at constant $F$ a
localization transition at temperature $T_c=1/2 m_2 F$.

All cases with $m_2 \neq 0$ discussed above have a common asymptotics
for large driving forces, which is determined by $m_2$ alone. The
reason therefore is, that this asymptotics is governed only by the
short-scale fluctuations of the potential, determined by leading term
of the stiffness for large $k$.

\subsection{Poissonian model}

Finally, we consider a model for diffusion on a stepped crystalline
surface, as illustrated in Fig.~\ref{capt_Poisson}a. This model shows
asymmetric transport properties and also exhibits a localization
transition.

The profile of the surface is described by the height function $h(x)$,
which increases in units of $a>0$ for increasing $x$. The structure is
characterized by the probability distribution $S(l)$ for the step
length $l$. We use a Poisson-distribution $S(l) = \exp(-l/l_0)/l_0$
with mean value $l_0$. The particle is supposed to be in constant
force-field with components $f_{x,h}$ in direction parallel and
perpendicular to the $x$-direction, which might be components of a
gravitational or electric field. They give rise to the total potential
$\phi(x)=f_h h(x)-f_x x$. In order to use our previous formulae, we
identify $U(x)=f_h h(x)-(a/l_0)f_h x$ and $F=f_x-(a/l_0)f_h$ by
subtracting the mean bias.

 From the Poissonian distributed number of steps in an interval of
length $y$ one obtains easily
\be{g.Poisson}
  g_T(y)=\exp[y F_{c1}^\pm (T)/T]
\ee
for $\xi \gtopl 0$ with asymmetric forces of first order
\be{Fc1.Poisson}
  F_{c1}^\pm (T)=-(a/l_0) f_h \mp (1- e^{\pm a f_h/T}) T/l_0 \gtopl 0 ,
\ee
which depend on temperature. From Eq.~(\ref{g.Poisson}) we obtain  (see
Fig.~\ref{capt_Poisson}b)
\be{v.Poisson}
  \mu_T(F)=\left\{ \begin{array}{ll}
  0 & {\rm for} \quad  F_{c1}^- (T) < F <F_{c1}^+ (T) ,\\
  1- F_{c1}^\pm(T)/F & {\rm for} \quad F \gtopl F_{c1}^\pm (T) .
  \end{array} \right.
\ee
Since at $T=0$ the localization region covers $-(a/l_0) f_h < F <
\infty$ (corresponding to $0 < f_{x} <\infty$) and shrinks to zero for
$T\rightarrow
\infty$, we find for forces in this region a localization transition at a
temperature $T_c$, which is implicitly determined by $F= F_{c1}^\pm (T)$, as
depicted in Fig.~\ref{capt_Poisson}c.

Apart from the asymmetry, the Poissonian model is very similar to the
random-force Gaussian model. Indeed, the force is distributed
independently at different sites, this time, however, with an
asymmetric and singular distribution.

\section{Conclusion}
\label{sec_5}

An analysis of particle mobility in one-dimensional disorder was
presented. We developed a framework based on the generating function
of spatial correlations of the disorder potential. For arbitrary
temperature and driving force, these correlations are relevant up to a
length $T/F$.  At large driving forces, mobility depends only on local
properties of disorder, whereas at small driving forces global aspects
matter.

For some models, which generalize previously studied structures of
disorder, mobility was evaluated over the complete range of
temperature and force. Thermally activated motion led to a rich
phenomenology.

The temperature-dependence of mobility can deviate drastically from
Arrhenius formula. This is characteristic for systems with a broad
distribution of activation energies. Such deviations have been
obtained already by an ad hoc averaging of escape times over a
spectrum of activation energies (see e.g.
\cite{Vil88}). This procedure is unsatisfying from a principal point of view,
since transport properties depend also on the spatial location and not only on
the height of energy barriers, which is in higher dimensions even more
important than in one dimension.

The models with spatially uncorrelated force distribution, the
Poissonian model as well as the random-force Gaussian model, exhibit
a localization transition. The similar phenomenology of these models
is due to the validity of the Central Limit Theorem, which assures a
Gaussian distribution of $U(x+y)-U(x)$.

It is therefore of particular interest to examine the stability of the
localization phases with respect to the introduction of correlations
in the force-distribution. For the Gaussian model, we achieved this by
two different, continuous deformations of the random-force model.
Switching on the couplings $m_0$ or $m_1$, we found the localized
phase to be {\em unstable}, i.e.  mobility was finite for all positive
temperatures and driving forces.

This result complements previous studies: In a dynamics, where the
particle may hop on a lattice only in one direction and which is thus
incompatible with a Langevin equation of motion, dynamical phases were
found to be stable with respect to short-ranged correlations in the
hopping rates, but unstable with respect to long-ranged
correlations\cite{Asl+.corr}. Introducing force-correlations into the
Sinai model, the scaling behavior of diffusion in the absence of a
driving fore has been found to be modified, but remained
anomalous\cite{Bou+87JP,Hav+89}.

\section{Acknowledgements}

I am grateful to P. Nozi\`eres for a critical reading of the
manuscript and to N. Pottier and D. Saint-James for interesting
comments.

\newpage
\figone
\figtwo
\newpage
\noindent
\figthree

\end{multicols}
\end{document}